\documentclass[submission, Phys]{SciPost}
\usepackage{amsmath,cases}
\usepackage[utf8]{inputenc}
\usepackage[T1]{fontenc}
\usepackage{lmodern}
\usepackage{graphicx}
\usepackage{amssymb}
\usepackage{gensymb}
\usepackage{bm}
\usepackage{tabularx}
\usepackage{mathtools}
\usepackage{microtype}
\usepackage{siunitx}
\usepackage[version=4]{mhchem}
\usepackage{bbm}
\usepackage[justification = raggedright]{caption}
\usepackage{braket}
\usepackage[normalem]{ulem}
\usepackage{layouts}
\usepackage{float}
\usepackage{textgreek}
\usepackage{orcidlink}
\usepackage{multirow}
\usepackage{amsfonts}
\usepackage{todonotes}
\usepackage[frozencache,cachedir=minted-cache]{minted}
\usepackage{framed}
\usepackage[super]{nth}
\usepackage{glossaries}
\usepackage{xspace}
\usepackage{hyperref}
\hypersetup{colorlinks}

\definecolor{bg}{rgb}{0.95,0.95,0.95}
\setminted{fontsize=\fontsize{9.5}{11}\selectfont, bgcolor=bg, baselinestretch=1, breaklines=true}
\newcommand{\comment}[1]{\stepcounter{CommentNumber}\belowpdfbookmark{#1}{\arabic{CommentNumber}}}

\newcommand{\ScatterWorks}{\texttt{ScatterWorks}\xspace}
\newcounter{CommentNumber}

\begin{document}

\begin{center}{\Large \textbf{
      ScatterWorks: A Python package for building and solving scattering network models
    }}\end{center}

\begin{center}
R.~Johanna~Zijderveld$^{1,\star}$,
H\'{e}l\`{e}ne Spring$^{1}$,
Anton~R.~Akhmerov$^{1,\dagger}$
\\[0.5em]
$^{1}$Kavli Institute of Nanoscience, Delft University of Technology, 2600 GA Delft, The Netherlands\\
${}^\star$ {\small \sf johanna@zijderveld.de}\quad ${}^\dagger$ {\small \sf scatterworks@antonakhmerov.org}
\end{center}


\begin{center}
  August 28, 2026
\end{center}

\section*{Abstract}
\textbf{
Network models provide an efficient framework for studying non-interacting transport and wave-propagation phenomena in disordered and topological systems.
We introduce \ScatterWorks, an open-source Python package for building and solving network models.
The package uses a compact network representation with composable transformation operations (\texttt{tile}, \texttt{cut}, \texttt{relink}, and \texttt{union}) to enable flexible construction of periodic and finite networks.
Local scattering matrices on nodes can be assigned by node labels or explicit indexing, assembled into global scattering equations, and used to compute quasienergies and transport observables.
The package supports sparse scattering matrices, allowing transport observables to be evaluated with Schur-based solvers.
We demonstrate the package workflow on a symbolic Fabry--Perot interferometer and a numerical Chalker--Coddington model, recovering analytical expressions and the expected near-critical transport behavior of the Quantum Hall transition.
Plotting utilities with label-aware rendering are included to streamline debugging and reproducible setup for larger, custom network geometries.
}

\tableofcontents

\section{Introduction: motivation}

\comment{Network models are a useful research tool for non-interacting scattering problems.}
Network models encode non-interacting transport in terms of wave amplitudes on links and local scattering relations at nodes.
They have been used to study criticality and scaling in disordered systems, for example in the Chalker-Coddington network model \cite{Chalker1988}.
Furthermore, they have also been applied to moiré systems, including twisted bilayer graphene \cite{Efimkin2018, DeBeule2021, Wittig2023, DeBeule2020, Chou2020, Vakhtel2022}, and in Kagome settings \cite{Wittig2024, Pal2019, Moulsdale2022}.
Network models have also been applied to strained graphene structures \cite{DeBeule2023}, skyrmionic lattices \cite{Wilczak2025}, and Floquet problems \cite{Delplace2020, Potter2020, Pasek2014}.
A comprehensive review on network models for different symmetry classes and generalizations of the Chalker-Coddington network model is given in~\cite{Kramer2005}.

\comment{On top of the places where network models are already used, there are many places where their use could be further explored.}
Despite their versatility, network models remain less widely used than continuum and tight-binding approaches in parts of the condensed matter community, which leaves many settings where their use could be further explored.
As a concrete example, transfer matrix methods are often used to calculate conductances, but in some settings they can suffer from numerical instabilities associated with the exponentially growing and decaying modes of the transfer matrices.
Moreover, many researchers rely on private, problem-specific codes that are difficult to repurpose for different network model applications.
As a consequence, changes in geometry, boundary type, or disorder protocol can require reworking core setup logic rather than only changing model parameters.

\comment{While many people have their own local codes for studying network models, this package offers a unified approach for studying more complicated networks.}
In order to aid the study of physical phenomena using network models, we have developed an open source python package named \ScatterWorks.
The \ScatterWorks package offers an advantage by providing a unified approach to working with network models.
We developed the package such that it can handle a range of nontrivial network model problems, including:
\begin{itemize}
\item the ability to create network models with symbolic scattering matrices to allow for analytical calculations;
\item the creation of complex unit cells, network ribbons, and closed geometries with boundary scattering matrices;
\item large networks, where we solve for blocks of the network and then recursively recombine them into smaller networks.
\end{itemize}

\comment{A second design principle is scalable, numerically efficient computation.}
In addition to the flexibility of network creation, the package is built so that construction and solution remain tractable for large networks.
\ScatterWorks uses sparse formulations throughout and supports MUMPS-based sparse linear algebra \cite{MUMPS:1, MUMPS:2} for core solves, which improves practical performance for large networks where dense methods become limiting.
This design enables larger parameter sweeps and more complex geometries without changing the modeling interface.

\comment{The manuscript is organized as follows.}
After introducing the basic network model formalism, we first show a Fabry-Perot network model which uses symbolic scattering matrices as an example.
We then show how a numerical Chalker-Coddington network model can be created and used to calculate the conductance near the critical point.
Finally, we discuss the technical details of the package, including the network representation, network operations, scattering equation compilation, and solver choices.

\section{The basics of network models}
\comment{A network model is defined as a collection of nodes and links, where the links carry a wavefunction amplitude and the nodes determine the scattering between links.}
A network model is a set of local scattering matrices connected by links.
Each link carries a single wavefunction amplitude, and each node relates the amplitudes of the incoming links to the amplitudes of the outgoing links.
As an example, if we have a node with incoming links $\psi_2,\psi_3$ and outgoing links $\psi_0,\psi_1$, then the local scattering relation at the node has the form:
\begin{equation} \label{eq:basicnode}
\begin{pmatrix}
\psi_0 \\
\psi_1
\end{pmatrix}
=
\begin{pmatrix}
s_{0,2} & s_{0,3} \\
s_{1,2} & s_{1,3}
\end{pmatrix}
\begin{pmatrix}
\psi_2 \\
\psi_3
\end{pmatrix}.
\end{equation}

\begin{figure}[bth]
  \centering
  \includegraphics[width=1\columnwidth]{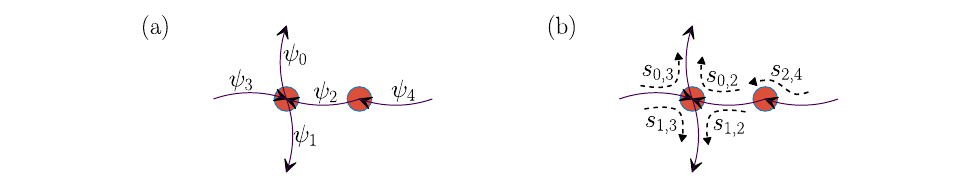}
  \caption{
    \label{fig:basic_network}
 Tiny basic network described by Eq.~\ref{eq:basicscatteringoperator}: (a) shows the link labels, and (b) indicates which scattering elements correspond to each link connection.
  }
\end{figure}

\comment{The local scattering matrices are inserted into a global sparse scattering matrix.}
In order to create a full network operator, we need to combine all the local scattering matrices into a single global matrix $S$ whose basis is the ordered list of all links.
For example, consider a five-link network with the node in Eq.~\ref{eq:basicnode} and a second one-channel node which scatters $\psi_4$ into $\psi_2$ with amplitude $s_{2,4}$.
This five-link network is shown in Fig.~\ref{fig:basic_network}.
Then the full link basis is $\Psi_{all}=(\psi_0,\psi_1,\psi_2,\psi_3,\psi_4)^T$ and the local scattering matrices are inserted into the global scattering matrix as
\begin{equation} \label{eq:basicscatteringoperator}
S =
\begin{pmatrix}
0 & 0 & s_{0,2} & s_{0,3} & 0  \\
0 & 0 & s_{1,2} & s_{1,3} & 0  \\
0 & 0 & 0 & 0 & s_{2,4}  \\
0 & 0 & 0 & 0 & 0 \\
0 & 0 & 0 & 0 & 0  \\
\end{pmatrix}.
\end{equation}
Additional nodes fill other rows and columns in the same way.
Thus, the matrix element $S_{\alpha,\beta}$ is nonzero when a node scatters amplitude from incoming link $\psi_\beta$ to outgoing link $\psi_\alpha$.
For a closed network with every local input and output port represented, current conservation at each node makes the fully assembled global scattering matrix unitary; the partially assembled matrix in Eq.~\ref{eq:basicscatteringoperator} does not yet satisfy these conditions.

\comment{This global scattering matrix describes the evolution of the full system, and can therefore be used for a quasienergy spectrum.}
This unitary scattering matrix, sometimes called the Ho-Chalker evolution operator, describes the evolution of the network after a single scattering step~\cite{Ho1996,Delplace2017}.
Therefore, its stationary states are wavefunctions that are unchanged up to a phase after one step:
\begin{equation} \label{eq:quasienergy}
S \Psi = e^{i E } \Psi,
\end{equation}
where $\Psi$ is a stationary state wavefunction and $E$ is the real eigenphase acquired in a single scattering step, defined modulo $2\pi$.
It becomes a quasienergy in the Floquet sense once a duration $T$ is assigned to one scattering step, and we use the term throughout with this identification understood.
For translationally invariant networks, the global scattering matrix $S$ depends on momentum $\bm{k}$ such that we get a periodic quasienergy band structure $E(\bm{k})$.

\comment{For transport properties, we first reorganize $S$ based on the connection to leads.}
Furthermore, $S$ also gives transport properties, such as the conductance, of a given network model~\cite{Beenakker1997, Fisher1981}.
To compute such transport, we need to first compute the effective scattering between the links which we select as external leads, where we distinguish between incoming lead links and outgoing lead links.
This distinction allows us to sort all links into three types of links: $\Psi_b$ which refers to the bulk links, $\Psi_i$ which refers to the incoming lead links and $\Psi_o$ which refers to the outgoing lead links.
When we make this distinction, the global scattering relation takes the block form
\begin{equation} \label{eq:hochalkerpartitioned}
\begin{pmatrix}
\Psi_b \\
\Psi_o
\end{pmatrix}
=
\begin{bmatrix}
S_{bb} & S_{bi} \\
S_{ob} & S_{oi}
\end{bmatrix}
\begin{pmatrix}
\Psi_b \\
\Psi_i
\end{pmatrix}.
\end{equation}
Here $S_{bb}$ scatters bulk links to bulk links, $S_{bi}$ scatters incoming lead amplitudes into bulk links, $S_{ob}$ scatters bulk amplitudes into outgoing leads, and $S_{oi}$ describes direct scattering from incoming to outgoing leads.

\comment{We can use this reordered scattering matrix to obtain the effective scattering matrix.}
To obtain the effective scattering, we aim to eliminate the bulk links and keep only the relation between the amplitudes entering and leaving through the leads.
For fixed incoming lead amplitudes $\Psi_i$, the first row of
Eq.~\ref{eq:hochalkerpartitioned} gives
\begin{equation} \label{eq:psib}
    \Psi_b = (I-S_{bb})^{-1}S_{bi}\Psi_i .
\end{equation}
The inverse $(I-S_{bb})^{-1}$ is defined whenever $S_{bb}$ has no eigenvalue equal to one.
If $S_{bb}$ has an eigenvalue equal to one, the bulk hosts a state that neither radiates into the outgoing leads nor is fed by the incoming ones.
Such a decoupled internal resonance makes $I-S_{bb}$ singular.
Substituting Eq.~\ref{eq:psib} into the expression for $\Psi_o$ gives
\begin{equation}
\begin{aligned}
    \Psi_o
    &= S_{ob}\Psi_b + S_{oi}\Psi_i \\
    &= \left[
        S_{ob}(I-S_{bb})^{-1}S_{bi} + S_{oi}
    \right]\Psi_i .
\end{aligned}
\end{equation}
Therefore, the effective scattering matrix between the incoming and outgoing
lead amplitudes is
\begin{equation}
    S_{\mathrm{eff}}
    =
    S_{ob}(I-S_{bb})^{-1}S_{bi} + S_{oi},
\end{equation}
where $I$ is the identity matrix on the bulk-link subspace.
For a two-terminal system, let $t$ denote the block of $S_{\mathrm{eff}}$ that transmits incoming channels from the left lead into outgoing channels in the right lead.
The zero-temperature linear-response conductance is then given by the Landauer formula
\begin{equation}
    G = \frac{e^2}{h}\operatorname{Tr}(t^\dagger t)
      \equiv \frac{e^2}{h}g,
\end{equation}
where $g$ is the dimensionless conductance~\cite{Landauer1957, Buttiker1986}.

\section{Symbolic Fabry-Perot interferometer} \label{sec:fabry}
\begin{figure}[bth]
  \centering
  \includegraphics[width=1\columnwidth]{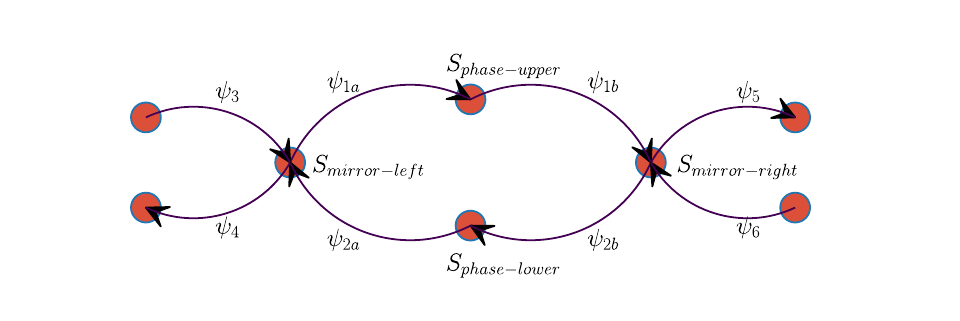}
  \caption{
    \label{fig:fabryperotsetup}
    The basic network setup with node and link labels of the Fabry-Perot interferometer.
    We omit the node labels of the boundary nodes at the sides, because they are only used for bookkeeping.
    }
\end{figure}

\comment{We use the Fabry-Perot interferometer as a minimal software-validation benchmark.}
In this section, we show how the \ScatterWorks package can be used to assemble a small symbolic network for a Fabry-Perot set-up and recover the known effective scattering matrix.
The optical setup has two partially transmitting mirrors and one incoming and outgoing channel~\cite{Perot1899}.
In the network representation, each mirror is a 2$\times$2 scattering node.
The links which traverse the cavity are split into two links so a 1$\times$1 phase node can be inserted and assigned a symbolic propagation phase.
Specifically, $\psi_1$ and $\psi_2$ are rewritten as $\psi_{1,\mathrm a}\to\psi_{1,\mathrm b}$ and $\psi_{2,\mathrm a}\to\psi_{2,\mathrm b}$.
Boundary nodes are used only to facilitate the existence of incoming/outgoing lead channels.
Throughout this manuscript, the listings assume the imports
\begin{minted}{python}
import numpy as np
import pandas as pd
import sympy

import scatterworks as sw
\end{minted}
so that \texttt{sw} exposes the whole public API.
To define the network we provide a NumPy array with the structure \texttt{[start\_node,end\_node]}:
\begin{minted}{python}
links_interferometer = np.array(
    [
        [0, 2],  # psi1a: mirror-left -> phase node
        [2, 1],  # psi1b: phase node -> mirror-right
        [1, 3],  # psi2a: mirror-right -> phase node
        [3, 0],  # psi2b: phase node -> mirror-left
        [0, 4],  # psi3: mirror-left -> left outgoing channel
        [5, 0],  # psi4: left incoming channel -> mirror-left
        [1, 6],  # psi5: mirror-right -> right outgoing channel
        [7, 1],  # psi6: right incoming channel -> mirror-right
    ],
    dtype=int,
)
network = sw.Network(
    links_interferometer,
    node_labels=pd.DataFrame(
        {
            "node_family": [
                "mirror",
                "mirror",
                "phase",
                "phase",
                "boundary_sink",
                "boundary_source",
                "boundary_sink",
                "boundary_source",
            ],
        },
    ),
)
\end{minted}
Here we also gave each node a label using the \texttt{node\_labels} attribute to facilitate assigning scattering matrices.

\comment{We define the scattering matrices of the mirrors to obtain the total scattering.}
The basis of the scattering matrices is defined by the order in which links first appear in the network.
Then the basis, and entries to the two mirror scattering nodes are:
\begin{equation}
S_{\text{mirror1}} =
\begin{pmatrix}
S_{1\mathrm a, 2\mathrm b} & S_{1\mathrm a, 4} \\
S_{3, 2\mathrm b} & S_{3, 4}
\end{pmatrix}
=
\begin{pmatrix}
r & t \\
t & -r
\end{pmatrix}
\!,
\end{equation}
and
\begin{equation}
S_{\text{mirror2}} =
\begin{pmatrix}
S_{2\mathrm a, 1\mathrm b} & S_{2\mathrm a, 6} \\
S_{5, 1\mathrm b} & S_{5, 6}
\end{pmatrix}
=
\begin{pmatrix}
r & t \\
t & -r
\end{pmatrix}
\!,
\end{equation}
where $t$ refers to the transmission of a mirror and $r$ refers to the reflection.
For real $r$ and $t$, this convention is unitary when $r^2+t^2=1$.
Both 1$\times$1 phase nodes in the cavity share the \texttt{phase} family,
\[
S_{\mathrm{phase}} = e^{ i\phi},
\]
where $\phi$ is the phase.
The boundary nodes are bookkeeping connectors:
\[
S_{\mathrm{boundary\_sink}}=\mathbf 0_{0\times 1},\qquad
S_{\mathrm{boundary\_source}}=\mathbf 0_{1\times 0}.
\]
To combine these scattering matrices into the full scattering equations, we use the previous node labelling and a dictionary structure:
\begin{minted}{python}
s_matrices = {
    "mirror": sympy.Matrix([[r, t], [t, -r]]),
    "phase": sympy.Matrix([[sympy.exp(sympy.I * phi)]]),
    "boundary_sink": sympy.zeros(0, 1),
    "boundary_source": sympy.zeros(1, 0),
}
S = sw.scattering_equations(
    network,
    s_matrices,
    by="node_family"
)
\end{minted}
We calculate the effective scattering matrix by defining the links which connect to the boundary nodes as leads.
We automatically identify these links and calculate the solution as follows:
\begin{minted}{python}
boundary_nodes = np.array([4, 5, 6, 7])
incoming_leads = np.nonzero(np.isin(network.links[:, 0], boundary_nodes))[0]
outgoing_leads = np.nonzero(np.isin(network.links[:, 1], boundary_nodes))[0]
solution, _, _ = sw.solve_scattering_equations(
    S,
    incoming_ind=incoming_leads,
    outgoing_ind=outgoing_leads,
)
\end{minted}
The solution then is:
\begin{equation}
S_{\mathrm eff}=
\begin{pmatrix}
\dfrac{r\left((r^2+t^2)e^{2 i\phi}-1\right)}{1-r^2 e^{2 i\phi}} & \dfrac{t^2 e^{i\phi}}{1-r^2 e^{2 i\phi}}\\
\dfrac{t^2 e^{i\phi}}{1-r^2 e^{2 i\phi}} & \dfrac{r\left((r^2+t^2)e^{2 i\phi}-1\right)}{1-r^2 e^{2 i\phi}}
\end{pmatrix},
\end{equation}
where the off-diagonal terms are transmission amplitudes and the diagonal terms are reflections.
Using the unitary mirror condition $r^2+t^2=1$, this simplifies to
\begin{equation}
S_{\mathrm eff}=
\begin{pmatrix}
\dfrac{r\left(e^{2 i\phi}-1\right)}{1-r^2 e^{2 i\phi}} & \dfrac{t^2 e^{i\phi}}{1-r^2 e^{2 i\phi}}\\
\dfrac{t^2 e^{i\phi}}{1-r^2 e^{2 i\phi}} & \dfrac{r\left(e^{2 i\phi}-1\right)}{1-r^2 e^{2 i\phi}}
\end{pmatrix},
\end{equation}
which is the standard Fabry-Perot transmission/reflection structure (denominator gives round-trip resonance), matching the literature~\cite{Ghamdi2023}.

\section{Numerical Chalker-Coddington network} \label{sec:cc}

\begin{figure}[bth]
  \centering
  \includegraphics[width=1\columnwidth]{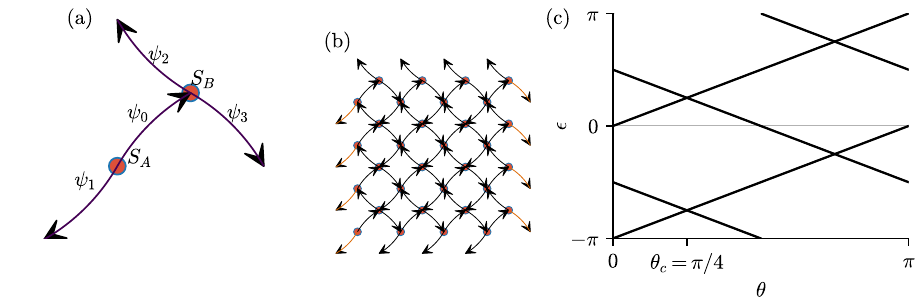}
  \caption{
    \label{fig:cc_energy}
    Panels (a) and (b) present the unit cell and tiled Chalker--Coddington network, respectively, while panel (c) gives the eigenphases of the unit-cell scattering operator at zero momentum as a function of $\theta$.
    The two eigenphase branches touch at $\theta=\pi/4$.
  }
\end{figure}
\comment{Most prominent example is a Chalker-Coddington network for the quantum hall phase.}
To demonstrate numerical performance of \ScatterWorks, we now consider the canonical Chalker-Coddington network~\cite{Chalker1988}.
This network describes the critical point of the quantum Hall phase transition~\cite{Zirnbauer1997}.
Each node in the network has two incoming links and two outgoing links, as is shown in Fig.~\ref{fig:cc_energy}(b).
The scattering matrix on each node is defined by:
\begin{equation} \label{eq:ccscatteringnodes}
S_A =
\begin{pmatrix}
S_{0, 2} & S_{0, 3} \\
S_{1, 2} & S_{1, 3}
\end{pmatrix}
=
\begin{pmatrix}
\cos \theta & i\sin \theta \\
i\sin \theta & \cos \theta
\end{pmatrix}
\!,
\end{equation}
and
\begin{equation}
S_B =
\begin{pmatrix}
S_{2, 0} & S_{2, 1} \\
S_{3, 0} & S_{3, 1}
\end{pmatrix}
=
\begin{pmatrix}
i\sin \theta & \cos \theta \\
\cos \theta & i\sin \theta
\end{pmatrix}
\!,
\end{equation}
where the first matrix once again shows the basis of the elements and the second matrix shows the scattering elements.
We constructed the scattering matrices such that scattering to the left always has probability $\sin^2 \theta$ while scattering to the right always has probability $\cos^2 \theta$.
This time, we want to define first only a unit cell of the network and then tile this unit cell into a bigger network.
We define the unit cell as two sublattices, labeled \texttt{A} and \texttt{B}, so that the two scattering matrices can be assigned by sublattice:
\begin{minted}{python}
cc_uc = sw.Network(
    np.array([[0, 1, 0, 0], #Psi_0
    [0, 1, -1, -1], #Psi_1
    [1, 0, 1, 0], #Psi_2
    [1, 0, 0, 1]]), #Psi_3
    node_labels=pd.DataFrame({
        "node_id": [0, 1],
        "sublattice": ["A", "B"],
        "x": [0.25, 0.75],
        "y": [0.25, 0.75],
    }),
    lattice_vectors=np.array([[1.0, 0.0], [0.0, 1.0]]),
)
\end{minted}
The final two columns in this NumPy array indicate whether a link crosses to a neighbouring unit cell.
Using this network and the scattering matrices in Eq.~\ref{eq:ccscatteringnodes}, we assemble the unit-cell scattering operator:
\begin{minted}{python}
s0 = np.array([[np.cos(theta), 1j * np.sin(theta)],
               [1j * np.sin(theta), np.cos(theta)]])
s1 = np.array([[1j * np.sin(theta), np.cos(theta)],
               [np.cos(theta), 1j * np.sin(theta)]])
s_matrices = {"A": s0, "B": s1}
S_full = sw.scattering_equations(
    cc_uc, s_matrices, by="sublattice", sparse=False
)
\end{minted}
The quasienergies are obtained from the phases of the eigenvalues of the unit-cell scattering operator at zero momentum, consistent with Eq.~\ref{eq:quasienergy}.
\begin{minted}{python}
eigvals = np.linalg.eigvals(S_full)
quasienergies = np.angle(eigvals)
\end{minted}
Figure~\ref{fig:cc_energy}(c) shows that the energy bands cross at $\theta = \pi/4$.
This is the critical point of the Chalker-Coddington model, because at this value the probability of scattering to the left is equal to the probability of scattering to the right~\cite{Chalker1988, Zirnbauer1997, Kramer2005}.
Far away from this critical $\theta_c$ the network is in a deeply localized phase, while near $\theta_c$ the localization length is determined by the symmetry class.

\begin{figure}[bth]
  \centering
  \includegraphics[width=1\columnwidth]{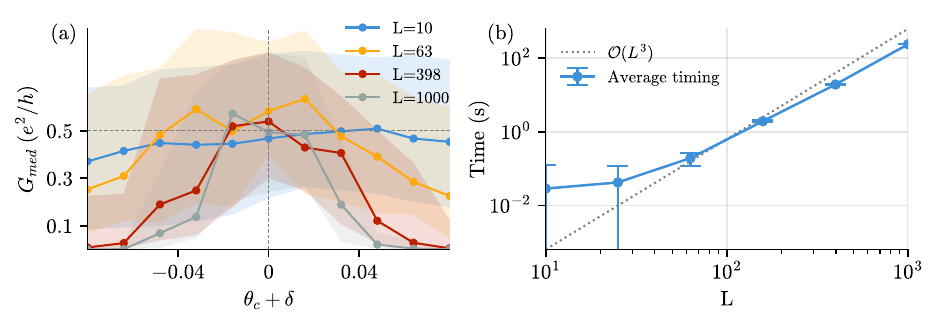}
  \caption{
    \label{fig:cc_conductance}
    Conductance and runtime scaling in the Chalker--Coddington model.
    Panel (a) plots the disorder-median conductance versus $\theta$ near $\theta_c=\pi/4$ for different system sizes $L$; the shaded regions span the 16th--84th percentiles across 50 disorder realizations.
    The conductance peaks as expected near $\theta=\pi/4$ at about $0.5\,e^2/h$.
    Panel (b) plots the mean runtime versus $L$ on log--log axes; the error bars show the standard deviation over 50 random seeds and all values of $\theta$, and the dotted $\mathcal{O}(L^3)$ line is a reference guide.
  }
\end{figure}
\comment{After construction of the links, the s-matrices get combined into the scattering equations and we can easily calculate numerical quantities such as the conductance.}
To calculate this conductance, we assign each physical directed link $\ell$ one independent random phase $\phi_\ell$, drawn uniformly from $[0,2\pi)$.
In contrast to the phase treatment in Sec.~\ref{sec:fabry}, we absorb the phase into the node scattering matrices by splitting it symmetrically between the source and sink of the link:
\begin{equation}
\begin{aligned}
S_n(\theta,\phi)
&=
D_{\mathrm{out},n}\,
S_{\sigma(n)}(\theta)\,
D_{\mathrm{in},n},
\qquad
\sigma(n)\in\{A,B\},
\\
D_{\mathrm{out},n}
&=
\operatorname{diag}\left(
e^{i\phi_{\ell^{\mathrm{out}}_{n,1}}/2},
e^{i\phi_{\ell^{\mathrm{out}}_{n,2}}/2}
\right),
\\
D_{\mathrm{in},n}
&=
\operatorname{diag}\left(
e^{i\phi_{\ell^{\mathrm{in}}_{n,1}}/2},
e^{i\phi_{\ell^{\mathrm{in}}_{n,2}}/2}
\right).
\end{aligned}
\end{equation}
Here $\ell^{\mathrm{out}}_{n,j}$ and $\ell^{\mathrm{in}}_{n,j}$ denote the outgoing and incoming links at node $n$, respectively.
The factors associated with the two endpoints combine so that propagation through link $\ell$ acquires the phase $e^{i\phi_\ell}$ exactly once.
To construct the scattering operator of the larger system, we first tile the unit-cell network:
\begin{minted}{python}
network = cc_uc.tile((length, length))
\end{minted}
After tiling, the resulting network remains periodic: links at the boundaries wrap around to form a closed network.
We then calculate the scattering operator of the full network:
\begin{minted}{python}
S_full = sw.scattering_equations(
    network, s_matrices_w_phase, by="sublattice", sparse=True
)
\end{minted}
Here \texttt{s\_matrices\_w\_phase} denotes a sublattice-keyed dictionary of callables, which assign each tiled node a scattering matrix with its own random phases.
For numerical efficiency, we construct this scattering operator in sparse format, which enables us to calculate the conductance for larger network sizes.
Before computing the conductance, we first determine the total scattering matrix between the external lead channels.
We define the leads as the links on the left and right boundaries of the network.
We can identify these by selecting the nonzero entries in the third column of \texttt{network.links}; these are the links that cross the horizontal periodic boundary.
After defining these links as external leads, we use the Schur solver to eliminate all remaining internal links and obtain the reduced scattering matrix between the leads.
\begin{minted}{python}
incoming_and_outgoing = np.nonzero(network.links[:, 2])[0]
reduced_scattering_matrix = sw.schur_solve(S_full, incoming_and_outgoing)
\end{minted}
This Schur solver uses the Schur complement to solve for the scattering~\cite{MUMPS:1, MUMPS:2}.
The reduced scattering matrix is expressed in the basis of the provided lead indices.
To know which boundaries the lead indices refer to, we consider that the \texttt{tile} operation preserves the link ordering of the unit cell.
Because the left-boundary link $\Psi_1$ is listed before the right-boundary link $\Psi_2$, the indices in \texttt{incoming\_and\_outgoing} are ordered by boundary, with the left-boundary channels followed by the right-boundary channels.
In this basis, the diagonal block of the reduced scattering matrix is the left-to-right transmission matrix, because its columns represent waves incident from the left and its rows represent waves leaving on the right.

\comment{We improve the computation time of large networks by using the Schur algorithm.}
Finally, we use this reduced scattering matrix to calculate the conductance~\cite{Landauer1957, Buttiker1986}:
\begin{minted}{python}
split = len(incoming_and_outgoing) // 2
transmission_matrix = reduced_scattering_matrix[split:, split:]
conductance_value = (np.abs(transmission_matrix) ** 2).sum()
\end{minted}
Figure~\ref{fig:cc_conductance} shows the result for multiple $\theta$ values near $\theta_c$ and shows that we numerically reproduce the expected behaviour.
For the finite $L\times L$ geometry considered here, with left and right leads and periodic transverse boundary conditions, the disorder-median conductance at $\theta_c$ is approximately $0.5\,e^2/h$.
Analogous calculations can be used to estimate the scaling exponent $\nu$, as demonstrated in ~\cite{MacKinnon1981, Kramer1993, Amado2011, Slevin2012}.
Because network-model calculations are particularly useful in regimes where the localization length can be very long, computational efficiency is an important requirement for \ScatterWorks.
This section has shown both how \ScatterWorks can be used with numerical scattering matrices and how sparse methods extend the accessible range of network sizes.
At the same time, the conductance calculation remains limited by the bottleneck step of the Schur solver, whose runtime follows the expected $\mathcal{O}(L^3)$ scaling with $L$ the linear system size~\cite{Collier2012}, as shown in Fig.~\ref{fig:cc_conductance}.

\section{Technical details}

In this section, we will discuss the technical details pertaining to the \ScatterWorks package.
We first explain the format, which includes the main objects which go into the package before showing the operations and default workflow which one might want to apply.

\subsection{Format}

\comment{A network is minimally defined by a collection of canonical links and a linkmap and stored in a \texttt{Network} object.}
In \ScatterWorks, each network is represented by a \texttt{Network} object, whose minimal description is the integer array \texttt{canonical\_links}.
The \texttt{canonical\_links} stores rows of the form $[\text{source}, \text{sink}, \text{shift}_0, \text{shift}_1, \dots, \text{shift}_{d-1}]$, where $d$ denotes the network dimension.
Each row specifies a directed link from a source node to a sink node, together with optional shift entries.
These shifts indicate the relative unit cell of the sink node: a zero shift connects nodes within the same unit cell, whereas a nonzero shift denotes a link from the source node in the reference cell to a sink node in a neighboring cell.
Periodic networks are useful both for momentum-space calculations, such as quasienergy spectra, and for constructing closed network geometries.
For purely finite networks, the shift columns can be omitted entirely.

\comment{The linkmap exists such that the canonical basis is preserved and remains compatible with scattering-matrix orderings.}
The order of the canonical links determines the basis of the node scattering matrices when \ScatterWorks assembles the global scattering operator.
If an operation that reconnects link endpoints also changed this ordering, the affected scattering matrices would have to be permuted to remain consistent with the new basis.
To avoid this additional transformation, network operations preserve \texttt{canonical\_links} and record changes in connectivity separately.
Each \texttt{Network} object stores a \texttt{link\_map} array relating the network's current links to \texttt{canonical\_links}.
The first column in \texttt{link\_map} selects the canonical row that supplies the source endpoint, and the second column selects the canonical row that supplies the sink endpoint and shifts.
If the user omits the \texttt{link\_map} , \ScatterWorks uses the identity map \texttt{[[0,0], [1,1], ...]}, so that every canonical link is considered in the network.

\comment{The \texttt{Network} object also allows for extra optional metadata.}
Furthermore, a \texttt{Network} can store optional metadata, including \texttt{node\_labels} and \texttt{link\_labels} dataframes.
These dataframes are useful for plotting links, selecting specific subsets of links, and assigning scattering matrices.
Where possible, network operations automatically propagate and extend these dataframes.
Users can also add their own columns to the dataframes at any time.
When present, \texttt{lattice\_\allowbreak vectors} define the geometric periodicity used for tiling and plotting.

\subsection{Workflow}

\comment{The workflow is to create a network, perform operations on the network, associate scattering matrices and then create scattering equations.}
The workflow in \ScatterWorks is:
\begin{enumerate}
\item define a base \texttt{Network},
\item modify the network with operations on the network object,
\item inspect/debug the resulting network with labels and plotting,
\item compile local scattering matrices into global operator $S$,
\item solve for observables (quasienergies, conductance, etc.).
\end{enumerate}
In this section, we describe each step in detail, including key subtleties.

\comment{When creating the network initially, you consciously choose between creating a periodic or finite network, as well as choosing the link order.}
\emph{Step 1 - Network creation.}
The first step is to create a network by specifying, at minimum, its directed links.
At construction time, the user must decide whether the network is finite or periodic and, in the periodic case, choose its dimension, because this choice is encoded by the number of shift columns and cannot be changed by subsequent network operations.
At this stage, the \texttt{Network} describes only the topology; the local scattering matrices are assigned later.
Although the scattering matrices have not yet been assigned, the number of incoming and outgoing links at each node already determines their required dimensions, while the ordering of those links fixes their basis.
The link ordering should therefore be chosen deliberately when constructing the network.

\begin{figure}[bth]
  \centering
  \includegraphics[width=1\columnwidth]{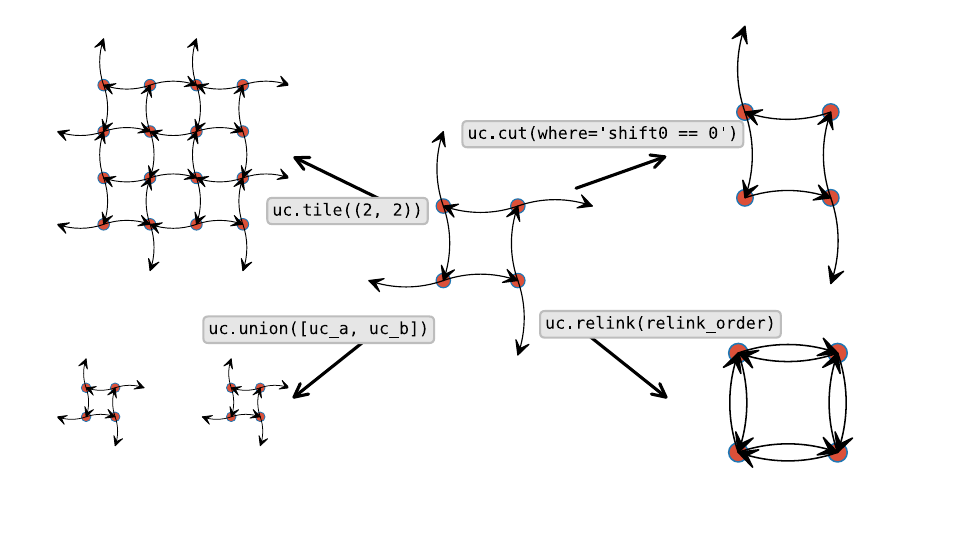}
  \caption{
    \label{fig:network_ops}
    The core \ScatterWorks operations applied to a simple unit cell: \texttt{tile}, \texttt{union}, \texttt{cut}, and \texttt{relink}.
    On the left, we apply \texttt{tile} and \texttt{union}; they increase the number of canonical links while preserving link order.
    On the right, we apply \texttt{cut} and \texttt{relink}; they modify \texttt{link\_map} while keeping \texttt{canonical\_links} fixed.
  }
\end{figure}

\comment{The tiling and union networkoperations extend \texttt{canonical\_links} while preserving the link order.}
\emph{Step 2 - Network operations.}
After creation of the network, we can apply different operations in order to manipulate the network into the desired form.
Figure~\ref{fig:network_ops} summarizes the main network operations available in the \ScatterWorks package.
\begin{itemize}
\item The \texttt{tile()} operation tiles the unit cell of a network and thereby increases the number of canonical links.
Furthermore, the \texttt{tile()} operation also scales lattice vectors to the supercell;
if coordinates are present in \texttt{node\_labels} (for example \texttt{x}, \texttt{y}), tiled coordinate columns are added without overwriting existing names.
\item \texttt{union()} returns a disjoint union of networks, shifts node ids of later blocks, and optionally adds a \texttt{network\_id} link label.
Therefore both the \texttt{tile()} and \texttt{union()} operation increase the number of canonical links in a network, while preserving link ordering.
\item We can use the \texttt{cut()} operation to continue with only a selection of links from the network.
For \texttt{cut()}, links can be selected by explicit IDs (\texttt{keep\_links}), node IDs (\texttt{keep\_nodes}), or by a dataframe query via \texttt{where=...}. The query is evaluated on \texttt{links\_df} with \texttt{join\_nodes=True}.
Internally, \texttt{cut()} modifies the \texttt{link\_map}, but keeps the \texttt{canonical\_links} fixed.
\item Similarly \texttt{relink()} also functions by changing \texttt{link\_map} for selected links.
When applying \texttt{relink()}, we reshuffle either sinks or sources.
For a relink operation, we provide the link numbers in a chosen order, and each selected link's source or sink is reconnected to the sink/source of the corresponding link in the sorted set of the same link numbers.
\end{itemize}
As a concrete example, when we do:
\begin{minted}{python}
network_relinked = network.relink(np.array([0, 3, 2]), mode="sinks")
\end{minted}
then the source of link $\psi_0$ is attached to the sink of link $\psi_0$, the source of link $\psi_2$ is attached to the sink of link $\psi_3$, and finally the source of link $\psi_3$ is attached to the sink of link $\psi_2$.
A key subtlety is that \texttt{relink()} changes connectivity by updating only \texttt{link\_map}, while each shift vector remains associated with its canonical sink endpoint.
If the shift inherited from the selected sink does not represent the desired unit-cell connection after relinking, it must be adjusted manually.

\comment{The node and link dataframes add explicit labels, which makes plotting-based network checks easier and filtering nodes or links more convenient.}
\emph{Step 3 - Checking with plotting and labelling}
In order to check that the network created after multiple network operations is as desired, it is frequently useful to inspect the resulting network by plotting it.
The plotting entry point \texttt{plot(...)} accepts a \texttt{Network} directly and supports hover/static label rendering for both nodes and links.
Instead of providing a network, we can also directly provide a list of x--y coordinates and links.
By default detached nodes are hidden (\texttt{show\_detached=False}), which is convenient after cuts on the network.
The label dataframes for both the link and the nodes provide not only a convenient way for selecting links to manipulate, but also provide a secondary checking mechanism when combined with plotting.
This is especially useful when the desired networks become more complicated and have local regions which differ from the remainder of the network.

\comment{The input of the scattering matrices has to be in the basis of the link order.}
\emph{Step 4 - Compiling scattering equations.}
After the desired network geometry has been achieved, we need to associate scattering matrices with the nodes to create the global scattering operator.
The function \texttt{scattering\_equations(network, s\_matrices, ...)} which we use for this, supports several input modes:
\begin{enumerate}
\item one global matrix or callable used for all connected nodes;
\item a dictionary keyed by values from a chosen node-label column. The \texttt{by} argument names that column;
\item an explicit assignment list whose entries have the form \texttt{(node\_\allowbreak ids, value)};
\item positional 3D batches, where each batch has shape \texttt{(n\_selected\_\allowbreak nodes, n\_out, n\_in)}.
\end{enumerate}
In all cases, dimensions must match each node's local link connectivity for the assigned scattering matrix.
With \texttt{by=...}, each group selected by one label value must have the same local in/out degree.
Only connected nodes require assignments; detached nodes are ignored.
Symbolic and numerical matrices (sparse and dense) are accepted, but symbolic and numerical objects cannot be mixed in the same function call.
To obtain the momentum dependency of a unit cell network, we need to explicitly call a separate function as follows:
\begin{minted}{python}
S = sw.scattering_equations(network, smatrices_by_family, by="family", sparse=True)
S_k = sw.k_dependency_scattering_equations(S, network, k=np.array([kx, ky]))
\end{minted}

\comment{After the scattering equations have been constructed, they can be solved between leads or used for other observables.}
\emph{Step 5 - Solvers, leads, and observables.}
To solve for the scattering between leads, both \texttt{solve\_scattering\_\allowbreak equations} and \texttt{schur\_solve} are available.
The difference between them is that \texttt{solve\_scattering\_\allowbreak equations} supports an arbitrary set of incoming and outgoing leads, while \texttt{schur\_solve} requires a single lead set that acts as both incoming and outgoing leads.
The \texttt{schur\_solve} method is preferable for large sparse systems, because it uses the Schur complement from the MUMPS library~~\cite{MUMPS:1, MUMPS:2}.
After scattering between a set of leads is calculated, it can be used to compute conductance or other observables:
\begin{minted}{python}
lead_idx = np.nonzero(network.links[:, 2])[0]
S_full = sw.scattering_equations(network, smatrices, sparse=True)
S_leads = sw.schur_solve(S_full, lead_idx)

t_block = S_leads[
    len(lead_idx) // 2:,
    len(lead_idx) // 2:,
]
G = (np.abs(t_block) ** 2).sum()
\end{minted}
The separation between network editing, scattering compilation, and lead solving is what makes the package practical for large and highly customized network constructions.

\section{Conclusion}

\comment{We showed the workflow of the \ScatterWorks package by going through a symbolic and numerical example.}
In this work, we introduced the \ScatterWorks package as a unified environment for building and solving network models across symbolic and numerical workflows.
The design is centered on a small number of composable steps: constructing a simple network, applying geometry operations to obtain the final network, assigning local scattering matrices, assembling global scattering equations, and computing observables.
We showcased this workflow with two examples: a symbolic Fabry-Perot interferometer, where the effective scattering structure is recovered analytically, and a numerical Chalker--Coddington model, where the expected quasienergy spectrum and conductance behavior near the critical point are reproduced.

\comment{The software package is flexible, efficient and tested.}
A key motivation for this software package is practical flexibility without sacrificing performance.
The implementation supports sparse formulations and the Schur-based transport routine enables calculations on larger networks than are practical with dense methods.
Node and link label dataframes enable assigning scattering matrices and selecting leads.
Furthermore, the plotting functionality, which is able to render hover or static labels, allows for direct visual inspection of the network structure, which is especially useful after applying complex geometry operations.
We also have tests on known results, from the closed form of the symbolic Fabry-Perot network in Sec.~\ref{sec:fabry} to the critical conductance $G \sim 0.5\,e^2/h$ in Fig.~\ref{fig:cc_conductance}.
The symbolic Fabry-Perot check runs in continuous integration on every commit, and together these checks provide a stable foundation for future development.

\comment{In the future the package will continue to improve and can be used in more situations.}
Looking forward, we expect to continue improving the user interface, in particular by easing the use of the relinking operation.
As \ScatterWorks is an open-source package, we welcome community contributions to expand functionality and improve ease of use.
More broadly, \ScatterWorks can be used to explore transport in other settings where network formulations are natural, including Floquet-driven physics~\cite{Delplace2020, Potter2020, Pasek2014} and random walk studies~\cite{Vakhtel2022}.
With these developments, network models can become a routine computational tool for transport studies beyond the examples presented here.

\section*{Acknowledgements}
We thank I.~Araya~Day for reviewing code during package development.
A.~A. and H.~S. were supported by NWO VIDI grant 016.Vidi.189.180 and by the Netherlands Organization for Scientific Research (NWO/OCW) as part of the Frontiers of Nanoscience program.

\section*{AI disclosure}
We have used generative AI tools to assist in both the code development and writing.
All AI contributions were reviewed and edited by R.~J.~Z. and A.~A., who take full responsibility for the content of the publication.

\section*{Data availability}
The code and full documentation for the \ScatterWorks package are available on Zenodo~\cite{ScatterWorks2026}.
The simulations and figures were produced with NumPy~\cite{Harris2020NumPy}, SciPy~\cite{Virtanen2020SciPy}, pandas~\cite{pandas}, SymPy~\cite{Meurer2017SymPy}, and Matplotlib~\cite{Hunter2007Matplotlib}.

\section*{Author contributions}
A.~A.~had the initial idea and oversaw the project.
H.~S.~contributed to early code development and package structuring.
R.~J.~Z. and A.~A. contributed to the package structure, code creation, and code review.
R.~J.~Z. wrote the paper.

\bibliography{bibliography.bib}

\end{document}